\documentstyle[12pt]{article}
\newcounter{num}
\begin{document}
\baselineskip 18pt
\def\lq{``}
\def\rq{''}

This article will appear in {\em Physics Essays}, Vol. {\bf 7}.

\noindent{\bf 1. Introduction}

Statistical Mechanics is a mature discipline.  As is the case in Quantum
Mechanics,$^{(1,2)}$ physicists know how to calculate any quantity of
interest, and no one doubts it will be in complete agreement with
experiment.  However, some physicists feel uneasy with the
interpretation of the rules of the game, because it appears to be in
conflict with the nearest and dearest part of their intuition.  This
Essay is concerned with the meaning of entropy; as evidence of the
efforts to reach a harmonious position in this issue, let me recall
Refs.~3-4.

The second law of thermodynamics rose as an empirical
fact and asserted that the useful energy available from an
isolated system will always decrease (and eventually approach zero). This
degradative behaviour is predicted not only for simple processes (such as
diffusion, heat conduction or explosions), but for every possible process, even
if methodic actions are taken by intelligent beings in the isolated system, no
matter how ingenious they might be. The only exception to this
degradative behaviour
are the so called ``reversible processes", in which useful
energy is kept unused.  In Section 2 I introduce ``the rebel'', and
attempt to analyze the reasons he has for not liking this form of the
second law.

With Lord Kelvin and Clausius, the second law became more formal.
Kelvin's formulation states that if a closed system is at uniform
temperature, then it is impossible to transform part of its internal
energy into mechanical energy.  Using the second law Clausius proved
that, for $T$ being the absolute temperature and $d' Q$ the heat
entering the system in an infinitesimal reversible process, $1/T$ is
an integrating factor for $d' Q$.  This result translates the second law
into the following assertions.

\noindent {\em a.} If a system is in equilibrium, there exists a
function of state $S$ (called the entropy), such that
\begin{eqnarray}
dS = d' Q/T \ .     
\end{eqnarray}
{\em b.} If a process occurs in an isolated system, the final entropy
will not be less than the initial entropy.

The next stage of evolution of the second law came in with Boltzmann's
$H$-theorem.$^{(5)}$  He built a function and showed that, under
apparently innocent assumptions, it behaves as the entropy.  Boltzmann's
theorem changed the status of the second law from an experimental fact
to a consequence of dynamics.  It also provided the previously abstract
function $S$ with a meaning in microscopic terms ($S$ could then be
interpreted as a measure of disorder).  The most controversial feature
of the $H$-theorem is the apparition of time-asymmetry in the
behaviour of a system which evolves under time-symmetric laws.  As a
hopefully representative short list of the never ending debate raised by
the $H$-theorem, I have picked Refs.~6-10.  Boltzmann's theorem will be
examined in Section 3.

Part of the controversial nature of the second law is probably due to
the usage of the same words by different authors, though they have
different concepts in mind.  In order to be on the safe side, I will
stick to the definitions in Section 3.  No claim is made that the
definitions given here are necessarily identical with those used by the
majority of the scientific community.  Hopefully, the adoption of a
non-conventional view will shed light and stimulate thought to a greater
degree than it will cause confusion.  Section 3 provides the foundations
for the discussion in Sections 5 and 6.

In Section 4 the rebel will be given the chance to beat the second law.
However,
his efforts fail.  In the remainder of this article, the rebel will have
to learn how to live with this law.

With the advent of information theory,$^{(11-15)}$ it became natural to
identify entropy with lack of information.  This is done in Section 5,
which is a corridor towards the closing section.  The connection between
information and entropy was hazy until Jaynes$^{(16-19)}$ and
followers$^{(20-22)}$ cast it into precise terms.

Section 6 contains what could be called the output of this article.  The
suggested conclusions are:

\noindent{\em a.} Disorder and lack of information are not the same
thing.\\
{\em b.} The notion of ``encoded order'' is introduced.\\
{\em c.} Quantities which are traditionally considered to be completely
objective, such as temperature, do depend on the awareness of the
thermometer designer about the state of the system.\\
{\em d.} The correct formulation of the second law should take ({\em c})
into account.

It is my conviction that the interpretation given here was already
implicit in the thinking of many physicists, such as J.C.
Maxwell,$^{(23)}$ A. Einstein, R.C. Tolman and E.T. Jaynes.  Perhaps
the
feature which distinguishes this article from all others is its attitude
towards ``the rebel''.  His obstinacy is not tagged as blindness to
facts, but is regarded as an expression of honest, courageous and
creative thinking.$^{(24)}$

The mathematical definition of ``randomness'' introduced by
Solomonoff,$^{(25)}$ Kolmogorov$^{(26)}$ and Chaitin$^{(27)}$ has led
several authors$^{(28,29)}$ to associate a larger entropy to
{\em microstates} which are ``less simple''.  This point will be
addressed in Section 6.

A more popular version of this article will appear elsewhere.$^{(30)}$

For the sake of completeness, I conclude this introduction by mentioning
related subjects which will not be treated here.

\noindent{\em a.} While
Schroedinger's equation is invariant under time-reversal, the
``collapse'' of the wave function is not.  This led von Neumann$^{(31)}$
to regard quantum measurements as  the ultimate reason for
irreversibility.  This aspect is given only indirect consideration in
Section 3.\\
{\em b.} In
principle, the second law applies to ``isolated systems''.  If we
define them as systems which do not interact with anything else, then
the only truly
isolated system would be the entire universe.  But if isolation implies
constant energy and volume, then the universe is an illegal example,
since it is expanding.  In addition, the universe is not a typical
thermodynamic system, since, as a consequence of the long range of
gravitation, the energy and the entropy are not extensive quantities.
Some references which consider the cosmologic aspects of thermodynamics
are 9 and 32-34.  The connection between the ideas in this article and
the expansion of the universe is discussed in Ref.~30.\\

\noindent
{\bf 2. The Rebel's Standpoint}
\par
The rebel does not respect the second law, which is an inequality rather than
an
equality. Since an inequality is more likely to be fulfilled by chance - as
compared to an equality, it is experimentally less convincing.
\par
The rebel has great self esteem, and does not accept the extrapolation that
mankind
will never be able to contrive in the future what couldn't be contrived in the
past.
When the rebel encounters a ``proof'' for the second law, his/her
conviction is that the only thing proven is the lack of imagination of
the prover.$^1$  The rebel never surrenders without a fight;$^{(24)}$
being a lone fighter only enhances his/her sense of mission.
\par
In spite of the long time elapsed since the universe began to evolve, the rebel
looks
around and discovers an amazing amount of order. Moreover, he/she is aware of
the
existence of ``historically irreversible" processes, such as the carving of a
bed by a
river, a ``favourable mutation" in Darwinian theory, or a discovery in a
communicating
society. After being  triggered by these processes, the produced changes build
up and
increase the amount of order perceived by their observer. All these phenomena
seem to
be opposed to the trend predicted by the second law.\\

\noindent
{\bf 3. Nomenclature and Boltzmann's Theorem}\\
{\bf 3.1 The statistical state and its evolution}

The definitions used here follow the line of thinking of Ref.~21.  I
shall use a quantum description of the system, not only because it is
the true representation of nature, but also because it is more
``natural'': it provides a natural way of counting states and assigning
them prior probabilities and a natural framework for a probabilistic
evolution.  Besides, it is more appealing to use sums rather than
integrals.
Nevertheless, I shall have in mind a
macroscopic system, so that  its behaviour in classical terms will also
be meaningful.
\par
Let $\psi _1,~\psi_2,...\psi_N$ be a basis of the Hilbert space of the possible
states of the system.  (If the system is isolated and has finite volume,
energy and degrees of freedom, then $N$ is finite too.)
{}From this paragraph on, whenever
I use the word ``state" (or ``microstate") without adding any ``family name", I
mean quantum state of the system.  Since the system is isolated, some
quantities,
such as the energy or the volume, remain constant. Only states with the
appropriate
value for those quantities (possibly within some tolerance) are possible. The
choice of the basis in the Hilbert space is arbitrary, and we usually select
the
basis which provides the simplest description to the best of our knowledge.
\par
By performing measurements, we can determine quantities such as the mass of the
upper
half of the system, the magnetization of the southern third of the system or
the
average pressure against its western wall. Such quantities are called
``macroscopic
coordinates" of the system. A set of macroscopic coordinates is called
``complete"
(within  a given set of constraints) if preparation of their initial values
determines reproducibly their evolution in time (allowing for the possible
exception of an initial transient time). Note that for the same given physical
system there may be several complete sets of macroscopic coordinates (some of
them
more detailed than others), depending on which quantities we are able or
willing to
measure. A complete set of values for the macroscopic coordinates is called the
``thermodynamic state" of the system.
\par
Generally, there will be a huge number of quantum states (i.e. microstates)
that
will fit the values of all the macroscopic coordinates in the complete set.
Therefore, knowledge of the thermodynamic state of a system does not determine
its
quantum state. As a consequence, the best we can do is to infer the
probabilities$^{2}$ $P_i$ for the system being in the state $\psi
_i~ (i=1,2,...N)$.
In the present discussion we disregard the possibility of the system
being in a
state $\sum c_i\psi_i$, with the coefficients $c_i$ having well defined phases.
This
policy is legitimate provided that the macroscopic coordinates are insensitive
to
these phases, either because they are diagonal in the basis $\{ \psi_i\}$ or
because
the contribution of the phases averages to zero.
The
set of probabilities $P_1,P_2,...P_N$ is
called here
the ``statistical state" of the system.  (The statistical state is
usually called the ``ensemble'' and the present policy about phases, the
``random phases assumption''.)

Is the statistical state ``objective'' or ``subjective''?  The precise
answer is that, for a given set of macroscopic coordinates and a given
basis of the Hilbert space (which are  chosen), the $P_i$'s are
{\em functions}$^{(16,20)}$ of the values obtained for the macroscopic
coordinates (which are measured).  The sorting of this situation is a
matter of semantics.

Let us now study how the statistical state evolves.  If at time $t_o$
the system were in a given state $\psi _i$, then at time
$t_o+\Delta t$ it would be in a linear combination of many of the states
$\psi_j$.
Since for complicated systems we cannot follow quantum phases, such linear
combinations look to us as mixtures, and our best description limits itself to
assigning a probability $W_{ij}$ for the passage from any initial $i$ to any
final
$j$. The values $W_{ij}$ depend on the dynamical details of the system and on
the
elapsed time $\Delta t$. It follows that if the statistical state at time $t_o$
was
$P_1,~P_2,~\ldots ~P_N,$ then the statistical state
$P_1^\prime,~P_2^\prime~\ldots
{}~P_N^\prime$ at time $t_o+\Delta t$ will be given by
\begin{eqnarray}
P^\prime _j=\sum_iW_{ij}P_i,              
\end{eqnarray}
where I have assumed Markovian evolution.$^{(35)}$
Clearly,
\begin{eqnarray}
\sum_jW_{ij}=1,          
\end{eqnarray}
since the system has to be in some state $j$ at time $t_o+\Delta t.$
In addition the linearity of Schroedinger's equation implies that every
$\psi_i$ will evolve into ${\cal U} \psi_i$, where ${\cal U}$ is a
linear operator.  Since the norms of wave functions do not change,
${\cal U}$ must be unitary.  Therefore, taking $W_{ij} = |(\psi_j|{\cal
U}\psi_i)|^2$, we also obtain
\begin{eqnarray}
\sum_iW_{ij}=1  \ . 
\end{eqnarray}

The set $\{P'_j\}$ in (2) is the most unbiased statistical state at time
$t_o+\Delta t$ which can be predicted from knowledge of the
thermodynamic state at time $t_o$.  This is not necessarily the same as
the statistical state that would be inferred if the complete set of
macroscopic coordinates were actually measured at $t_o+\Delta t$.  We
could predict
 $\{P'_j\}$ with certainty only if we could predict with certainty
the thermodynamic state.
It is tempting to argue that for a macroscopic system
and for $\Delta t$ negligible compared to Poincar\'{e}'s recursion time,
the variances in
 $\{P'_j\}$ will be very small and prediction (2) will be highly
reliable.  However, by definition, this argument is unnecessary: if the
macroscopic coordinates (and therefore
 $\{P'_j\}$) after an experimentally relevant time $\Delta t$ were not
determined uniquely by their initial values at $t_o$, then they would
not constitute a complete set.$^3$

The evolution rule (2) will be refined in Subsection 3.3.\\

\noindent{\bf 3.2 The entropy and its evolution}

Since it seems that no definition of entropy is universally accepted,
except for Eq.~(1), let me specify the choice of this article.  The
entropy will be temporarily given by
\setcounter{num}{5}
\setcounter{equation}{0}
\def\theequation{\thenum\alph{equation}}
\begin{eqnarray}
S = - k H           
\end{eqnarray}
with $k$ Boltzmann's constant and
\begin{eqnarray}
H = \sum_iP_i\ln P_i \ .         
\end{eqnarray}
This definition was used by Klein.$^{(36)}$  It will be revised in the
next subsection.
$H$ can be written in the form
\begin{eqnarray}
H = {\rm Tr} (\rho \ln \rho),          
\end{eqnarray}
where $\rho$
the reduced density matrix.
The reduced density matrix is obtained from the density matrix by
substituting the non-diagonal elements by zeroes.  An ugly
feature of (5c) is that $H$ depends on
 the particular basis $\psi_1, \psi_2, \dots \psi_N$
which we  choose.  It should become apparent below that this
ugliness is not a matter of mathematical clumsiness, but is an essential
feature of the $H$-function.

Definition (5) may be considered legal only if it reduces to definition
(1) when the latter is applicable, namely, infinitesimally close to a
situation in which the temperature is an equilibrium property and
energies can be attributed to the possible states of the system. This
implies the canonical distribution
\setcounter{equation}{5}
\def\theequation{\arabic{equation}}
\begin{eqnarray}
\ln P_i = - E_i/kT-\ln Z \ , 
\end{eqnarray}
where $E_i$ is the  energy of the system if it is at state
$\psi_i$ and the partition function $Z$ is independent of $i$.
Identifying
the internal energy with the expectation of $E$, and the work on the
system as the expectation of the change in $E$, we obtain
\begin{eqnarray}
d' Q = d (\sum_iP_iE_i) - \sum_iP_idE_i=\sum_iE_idP_i \ .    
\end{eqnarray}
On the other hand, using (5), (6) and noting that $\sum_i P_i$ does not
change
\begin{eqnarray}
TdS = - kT\sum_i(1+\ln P_i)d P_i = \sum_iE_id P_i \ ,   
\end{eqnarray}
in agreement with (1).

Definition (5) is not only a generalization of (1); it also has a
meaning: $H$ is the unique reasonable additive measure of  the
information$^{(11,12,21)}$ about the state of the system.

Let us now show that the entropy defined by (5) does not decrease with
time.  The following lines are a translation of the standard
proof$^{(8,37)}$ to present notation.  The evolution rule (2) gives
\begin{eqnarray}
H(t_o +\Delta t)=\sum_jP^\prime_j\ln P_j^\prime=
\sum_{ij}W_{ij}P_i\ln P_j^\prime          \   ,    
\end{eqnarray}
whereas $H(t_o)$ can be written in the form
\begin{eqnarray}
H(t_o)=\sum_i(\sum_jW_{ij})P_i\ln P_i=\sum _{ij}W_{ij}P_i\ln P_i.  
\end{eqnarray}
Therefore, since for any pair $(x,y)$ of positive numbers $x(\ln x - \ln
y)\geq x - y$, it follows from here and (4) that
\begin{eqnarray}
&&H(t_o)-H(t_o+\Delta t)=\sum_{ij}W_{ij} P_i (\ln P_i-\ln
P_j^\prime) \geq\nonumber \\
&&\sum _{ij} W_{ij}P_i-\sum_{ij} W_{ij}P_j^\prime =\sum_jP^\prime  
_j-\sum_jP_j^\prime =0 \ ,
\end{eqnarray}
as required by the second law.

Since Boltzmann was the first to build a function of the statistical
state of the system, such that it decreases as equilbrium is being
approached, inequality (11) will be called Boltzmann's theorem.
However, let me emphasize that definition (5b) is not equivalent to
Boltzmann's $H$.  A nonessential difference is that Boltzmann's $H$ was
classical.  An essential difference is that, whereas the index $i$ in
(5b) refers to the state of the entire system, Boltzmann's $H$ was
expressed in terms of the single-particle distribution function, and
could not take correlations into account.  As made clear by
Jaynes$^{(38,39)}$, Boltzmann's $H$ does not necessarily fulfill
inequality (11) for a system of interacting particles.\\

\noindent {\bf 3.3 Validity and significance of Boltzmann's theorem}

Let us first recognize that the range of application of the second law
has been greatly enlarged, from situations in which the system is in
equilibrium both at $t_o$ and at $t_o+\Delta t$ and a reversible path
connects between them, to any situation in which a given complete set of
macroscopic coordinates has well defined values at $t_o$
and, as a consequence, $\{P_i\}$ and
$\{P_i'\}$ are well defined, too.  Nevertheless, let us note that
definition (5) is only a possible generalization of (1) and, in the
absence of additional criteria, other generalizations may be possible,
too.

Next, let us note that an essential ingredient in the proof of (11) is
the assumption of
a probabilistic rather than a deterministic evolution of the quantum state.
If we knew that every state $\psi_i$ evolves during time $\Delta t$ into the
state
$\psi_{j(i)}$, then $W_{ij}$ would vanish for $j\ne j(i)$, equation
(2) would become
$P^\prime _{j(i)}=P_i,$ and $W_{ij}(\ln P_i-\ln P_j^\prime)$ would become
zero.     The absence of determinism may be traced to the fact that
measurement of the thermodynamic state is insensitive to phases;
therefore, the thermodynamic evolution can only be described by
averaging out the actual phases, and retaining only the probabilities
for the possible outcomes.

Indeed, had we taken $\rho$ as the full density matrix in definition
(5c), $H$ would remain constant in time due to the unitarity of the
operator ${\cal U}$.  This is most disturbing, since, with standard
modifications of (6)-(8), the expressions for $d' Q$ and for $TdS$ would
become \begin{eqnarray}
d' Q = {\rm Tr}({\cal H} d \rho)     
\end{eqnarray}
(where ${\cal H}$ is the Hamiltonian) and (5) would still be a
generalization of (1).  Let us now consider the case that the system is
in equilibrium at $t_o+\Delta t$.  How can two expressions for the
entropy both reduce to (1) at $t_o+\Delta t$ and nevertheless have
different values at $t_o+\Delta t$?  The answer is that expressions (7)
and (12) for $d' Q$ are different.

If so, we cannot decide which is the correct generalization for (1)
before we decide on the correct expression for $d' Q$.  The old
specification prescribed by thermodynamics is that heat is the amount of
energy transfer which cannot be accounted for as work, whereas work is
the change in mechanical energy involved in the changes of the
macroscopic coordinates.  Therefore, an expression for the work which
involves changes in quantum phases, while none of the macroscopic
coordinates is sensitive to them, is not thermodynamically acceptable.
Eq.~(12) must thus be rejected.

After learning this lesson, we should revise our definition of entropy.
For any realistic situation, the macroscopic coordinates will not only
be insensitive to the phases.  There will be classes of states $G_1, \
G_2, \dots G_M, \ M << N$, such that if $\psi_i$ and $\psi_j$ are both
in the same class $G_\nu$, then they will be experimentally
indistinguishable by measurement of any of the macroscopic coordinates.
Therefore, in the thermodynamic definition (7) of heat we should replace
the separate values $P_i$ for each state by common values within every
class.  At measurement time $t_o$, due to the complete inability to
distinguish among the states in $G_\nu$, all the $P_i$'s for $\psi_i\in
G_\nu$ will be equal anyway.  However, after time $\Delta t$, the
evolution rule (2) predicts in general different values $P'_j$ for
states $j$ in the same class.  These differences are thermodynamically
irrelevant.  The only thermodynamically relevant probabilities are
those of the entire classes, which equal the respective sums of the
$P'_j$'s pertaining to them.  If the system behaves reproducibly between
$t_o$ and $t_o+\Delta t$, measurement of the macroscopic coordinates at
$t_o+\Delta t$ should indeed give results which are compatible with the
predicted probabilities for the classes.  However, the least biased
probabilities to be associated with every microstate would not be the
$\{P'_j\}$, but rather the probabilities
$\{\overline{P'_j}\}$ which, while being compatible with the
probabilities of the classes, provide the smallest possible information
about the actual microstate of the system.  Since the smallest
information is available when all the probabilities within any given
class are equal,
\begin{eqnarray}
\overline{P'_j} = \ {\rm average \ of \ the} \
\{P'_j\} \ {\rm in \ the \ class \ of \ } j \ .       
\end{eqnarray}
This averaging of the evolved statistical state predicted by (2) is
usually called ``coarse graining''.$^{(8)}$ According to the discussion
here, a better name would be ``thermodynamic relevance''.
$^{(22)}$

It follows from the previous paragraph that the relevant entropy at
$t_o+\Delta t$ is not given by the $H(t_o+\Delta t)$ which is obtained
directly from (2).  Rather, we should use the value $\bar{H}(t_o+\Delta
t)$ which is obtained by substituting
$\{P'_j\} \rightarrow
\{\overline{P'_j}\}$ into the expression (5b) for $H$.  From (11) and
from the construction of (13)
\begin{eqnarray}
\bar{H}(t_o) = H(t_o) \geq H (t_o+\Delta t) \geq \bar{H}(t_o+\Delta t) \
, 
\end{eqnarray}
so that the entropy increases {\em a fortiori}.  The meaning of
inequality (14)
is that, if the thermodynamic state of an isolated system is known at time
$t_o$, it will evolve during the time interval $\Delta t$ in a way that implies
losing information about the actual microstate. As a consequence, we will lose
part
of our capability of utilizing the energy of the system.

If the states $\psi_i$ are good approximations to the eigenstates of
the Hamiltonian, then $W_{ij} = W_{ji}$, due to Fermi's golden rule.
This
condition is stronger than Eq.~(4) and is not always valid.$^{(40)}$
This stronger condition was assumed in Pauli's master equation$^{(41)}$
to prove a stronger inequality than (14), namely, $dH/dt \leq 0$.  For a
hierarchical analysis of $H$-theorems, see Ref.~42.

I shall close this subsection with the answers to two perennial
questions:\\
\noindent   {\em a.} Is the entropy objective or subjective?

Since the entropy is a function of the statistical state, both are
objective to the same extent.  Thus, the answer was already given in
Subsection 3.1.\\ {\em b}.  If we retrodict the statistical state at
$t_o-\Delta
t$ from the statistical state at $t_o$, Eqs.~(2) and (4) remain valid.
{}From them, we can use the same arguments that lead to (14) to prove that
$\bar{H}(t_o-\Delta t) \leq \bar{H}(t_o)$.  Actually, if we invoke the
invariance of Schroedinger's equation under time reversal, we reach a
stronger conclusion: the statistical state at $t_o-\Delta t$ is the same
as at $t_o+\Delta t$.  Is it true that the entropy decreases during the
time interval $(t_o-\Delta t,t_o)$?

If it were true that
the
only information we have about the state of the system comes from
the measurements
performed at $t_o$,
 and we had no idea about its
state in the past, then the answer to this question would be ``yes''.
This is not the usual experimental situation. In order to have a
meaningful
experiment, we have to prepare its initial thermodynamic state.
This is done either
by waiting for equilibrium and then removing some constraint,
or by following the
thermodynamic evolution until a definite regularity is detected.
This implies
knowledge about the system during some lapse of time before $t_o$.
The asymmetry  between past and future comes from the fact that
we usually
gather data about the past of the system, but we cannot gather
data about its
future. This question is treated in more detail in Refs.~30, 34, 43 and
44.\\

\noindent {\bf 3.4 Non-isolated systems}

Thermodynamics does not forbid the decrease of entropy for a system
which is not isolated.  Where in the proof of Eq.~(14) have we required
isolation of the system?

First, I have assumed that the Hilbert space remains unchanged
($\{\psi_i\}$ is still a basis at $t_o+\Delta t$).  This would not be
the case if the volume, or the shape, or the energy, or the Hamiltonian,
or the degrees of freedom of the system were to change.  Note, however,
that definition (5) permits an obvious generalization: for a
``microscopically adiabatic'' process, such that the Hilbert space or
the energy levels change, but no jumps among the states $\{i\}$ are
induced, the statistical state and the entropy do not undergo any
immediate change.

Second, I have assumed that, in principle, pure states remain pure.
Namely, the state of the system does not become entangled with the state
of some other system.  Entanglement would not affect the evolution rule
(2), but it would affect its interpretation.

Finally, I have used Eq.~(4).  In order to appreciate Eq.~(4), let us
consider a system which is not isolated and see that Eq.~(4) turns into
something else.

Let us examine the case of a
system in thermal contact with its environment.  By this we mean
a system which has to be considered together with a companion system,
called the
``environment'' in order to constitute  an
isolated system. Let their total energy be $E_T $ (possible
unknown). The coupling between
them is sufficiently weak, so that we can associate separate states to the
system
and
to the environment, with respective energies $E$ and $E_T-E$. On the other
hand,
this
coupling is sufficiently strong to bring about changes in $E$. For every value
of
$E$,
let there be $\omega (E)$ possible states for the system, which we shall denote
by
$i,~ 1\leq i \leq \omega (E)$. Similarly, the state of the environment will be
denoted
by $I, ~1\leq I \leq \Omega (E_T-E)$. We shall attempt to keep track of the
state of
the system (in a statistical sense); on the other hand, we shall have no
information
either about the state of the environment or about any possible correlation
between
$i$ and $I$.
\par
The state of the entire system will be determined by the three indices $E,i$
and $I$,
but we keep track only of the first two. The probability for a state $E,i$ of
the
system will be denoted by
$P_{Ei}=P_EP_i(E)$
where $P_i(E)$ is the probability of the state $E,i$, given the fact that its
energy
is $E$. Due  to our complete ignorance about the state of
the environment,
the probability $P_I(E)$ for a state $I$ when $E$ is known is simply
\begin{eqnarray}
P_I(E)=1/\Omega (E_T-E) \ .   
\end{eqnarray}
{}From the absence of correlation it follows for the state of the entire
system
\begin{eqnarray}
P_{EiI} = P_{Ei}/\Omega(E_T-E) \ .      
\end{eqnarray}

Let us now consider the probability $W_{Ei;E^\prime i^\prime}$ for passage of
the
system from $E,i$ to $E^\prime , i^\prime$. We wish to express $W_{Ei; E^\prime
i^\prime}$ in terms of $W_{EiI;E^\prime i^\prime I^\prime}$ - the probability
for
passage of the entire system from $E,i,I$ to $E^\prime, i^\prime, I^\prime$.
Since the
environment might initially be in any state $I$, and must finally be in some
state
$I^\prime$, we have
\begin{eqnarray}
W_{Ei;E^\prime i^\prime}=\sum_{II^\prime}P_I(E)W_{EiI;E^\prime i^\prime
I^\prime}={1\over \Omega (E_T-E)}\sum_{II^\prime}W_{EiI;E^\prime i^\prime
I^\prime}    \ .         
\end{eqnarray}
Since the entire system is isolated, we can apply Eq. (4) to $W_{EiI;E^\prime
i^\prime I^\prime}$. Using (17), it follows that
\begin{eqnarray}
\sum_{Ei}W_{Ei;E^\prime i^\prime}\Omega
(E_T-E)=\sum_{EiII^\prime} W_{EiI;E^\prime i^\prime
I^\prime}=\sum_{I^\prime}1=\Omega (E_T-E^\prime ) \ .   
\end{eqnarray}
  This is the relationship we were looking for: if a
system is in thermal contact with
its environment, Eq.~(4) has to be replaced by Eq.~(18).  A simple
meaning of Eq.~(18) is that a statistical state $P_{Ei} \propto
\Omega(E_T-E)$ is stationary.

Let us now develop a version of Boltzmann's theorem, adapted to the
present system.  Using (5b), (16), and summing over $I$, we obtain the
$H$-function for the system + environment:
\begin{eqnarray}
H_T=\sum_{Ei}P_{Ei}\ln {P_{Ei}\over \Omega (E_T-E)}  \ . 
\end{eqnarray}
Although $H_T$ is the value of $H$ for the system +
environment, it is
expressed in (19) in terms of the statistical state of the system only.
This result reflects the fact that $H$ measures information and we have
no information about the state of the environment.  Since the entire
system is isolated, we know already that $H_T(t_o+\Delta t) \leq
H_T(t_o)$.  It is an instructive exercise to show the decrease in $H_T$
by following the evolution of $\{P_{Ei}\}$.  This is done by the same
steps which lead to inequality (11), but condition (4) has to be
replaced by (18).

Eq.~(18) has the unsatisfactory feature that it involves the structure
of the environment (the function $\Omega$).  We can get rid of it when
the environment is a ``reservoir'', whose temperature is always $T$.
{}From (5) and (15),
\begin{eqnarray}
\Omega(E_T-E)=e^{S_{\rm res}/k} \ , 
\end{eqnarray}
where $S_{\rm res}$ is the entropy of the reservoir when the system has
energy $E$.  The temperature being constant and the work being zero, it
follows from definition (1) that
\begin{eqnarray}
S_{\rm res} = S_o - E/T  \ ,             
\end{eqnarray}
where $S_o$ is the entropy of the
reservoir when the energy of the
system vanishes.  Introducing (20) and (21), condition (18) reduces to
\begin{eqnarray}
\sum_{Ei}W_{Ei;E^\prime i^\prime }e^{-E/kT}=e^{-E^\prime /kT} \ . 
\end{eqnarray}

When the environment is a reservoir, $H_T$ has a simple thermodynamic
meaning.  Using (20) and (21), and dropping
an additive constant, $H_T$ reduces to
$<E>/kT-S/k$,
where $S$ is the entropy of the system and $<E>$ is the expectation
for its energy.  Therefore, $kTH_T$ is the Helmholtz free energy, which
is the appropriate thermodynamic potential for  a system kept at
constant temperature while no work is done on it.\\

\noindent{\bf 4. \ Attempts to Beat the Second Law}

The previous section proves that the entropy of an isolated system will
not decrease.  But you never know; perhaps the proof, and the massive
experimental evidence too, overlook some critical ingredient which might
do the job.  Therefore, let us be guided by the rebel's ingenuity and
devote one section to examine situations where this kind of ingredient
appears to be present.\\

\noindent
{\bf 4.1  Maxwellian Demons}
\par
The best known challenge to the second law was posed by J.C. Maxwell in 1871.
He
considered a volume of gas, initially at uniform temperature, divided into two
chambers by a partition with a door. Whenever a molecule attempts to go
through, a
``demon" decides whether to open or to close the door, depending on the speed
of the
molecule. In this way, the demon can manage to build a high concentration of
fast
molecules  in one chamber and slow molecules  in
the other. This would mean a decrease in the entropy of the gas. To be sure,
there is
nothing sacred about the speed of the molecules. As long as the demon can
detect some
feature of a molecule and behave selectively depending on whether it attempts
to
cross from chamber one to two or from two to one, a difference between the
chambers
will build up and this difference permits the extraction of useful energy from
the
gas.
\par
In order to pose a real challenge to the second law, the mysterious demon of
the
previous paragraph has to be replaced by a conceivable physical device. This
physical
``demon" should be some sort of unidirectional valve as encountered, for
instance, in
most inflatable objects. Two idealized candidates for this job are shown in
figures 1 and 2.
\par
Figure 1 shows a volume of gas, initially at uniform pressure, divided into two
chambers by a partition with a door which can open to one side only (say, to
the
left). A molecule (such as $M$) coming from the right will open the door and
cross
to the other chamber; the inverse process $(M^\prime )$ is inhibited. As a
consequence, molecules should accumulate at the left chamber and the pressure
difference could be used to perform work.  This demon, and his exorcism,
were proposed by Smoluchowski.$^{(45)}$
Situations which are apparently equivalent to that of
Fig.~1, though they are less transparent, are studied in Refs.~46 and
47.
\par
Figure 2 shows a resistor $R$ held at a positive absolute temperature. Due to
thermal motion of the electrons, a random voltage builds up between the ends of
$R$ (Johnson noise). This voltage cannot be used, because it changes sign
with an unpredictable rhythm. However, if we add a diode $D$ in series, we
should
obtain a rectified current $I$ which always circulates counterclockwise, and
could be
used to run motor $M$.  This possibility has been considered by several
authors.$^{(48)}$
\par
Devices of this sort do indeed work... provided that the physical
``demon" is
sufficiently cold. There is no particular novelty in these devices: they are
just
heat engines whose heat intake is at the gas (or at the resistor $R$), their
heat
sink is at the ``demon", and the difference in heat flows is used to perform
work.
They should not violate the second law more than any other existing heat
engine.
\par
The flaw in these devices (from the rebel's point of view) is that the ``demon"
heats up during operation. When the ``demon" is hot, it suffers from thermal
random movements which cause ``him" to perform an imperfect task. For instance,
the
door in Fig.~1 will begin to open and close by itself, even though no molecule
is hitting it. It may so happen that the door is open as shown in the figure,
the
spring is closing it, and a molecule is tracing a trajectory inverse to that of
$M$. Then this molecule would be kicked by the door into the right chamber
(that
is, into ``the wrong" chamber), in spite of the fact that it had no plans of
going there.
\par
Similarly, for positive absolute temperatures, the diode $D$ not only acts as a
rectifier, but also as a d.c. bias in ``the wrong" direction.
We may therefore expect, as it happens for
every heat engine, that the device will stop working at the moment that the
temperature of the ``demon" becomes equal to the temperature of the heat
source.
If the demon's temperature becomes larger than that of the ``heat source", then
the device should operate in ``the wrong'' direction.
\par
The rebel could rightly argue that the previous paragraph is a plausibility
argument
rather than a proof. The imperfections of the demon should
diminish his performance, but, before we see an explicit
calculation, why should we believe that the demon's action
will precisely cancel?  Moreover, when calculating the work performed by
the demon, we should avoid using the rules of statistical mechanics, or
even the concept of temperature, which already have the second law of
thermodynamics built into them; only Newton's laws (or Schroedinger's
equation) and mathematics are acceptable.  Calculations of this sort
have been recently performed:$^{(49)}$ simulations show indeed that
while the door in Fig.~1 is able to act as a unidirectional valve when
large pressure differences between the chambers are present, it does not
rectify fluctuations unless it is artificially cooled.

Smoluchowski concluded that ``there is no automatic, permanently
effective perpetual motion machine, in spite of molecular fluctuations,
but such a device might perhaps function regularly if it were operated
by intelligent beings.'' ``Intelligent beings'' differ from ``automatic
Maxwellian demons'' in that they have to take measurements and make
decisions.  These subjects will be dealt with in Section 5.

For a review of the literature  on Maxwellian demons, see Refs.~50 and
51.\\

\noindent{\bf 4.2 Geometry-based guides}

If you can't beat them -- join them.  In this subsection we consider
situations in which it is not attempted to attain rectification by
granting a preferred direction to the molecules, but, on the contrary,
by removing some preferred direction they already had.  The role of the
demon, who is merely the walls of the container, is just to make the
impinging molecules ``forget'' their original directions.  Therefore,
the more the walls quiver due to thermal agitation, the better their
performance.

As the simplest example consider a long thin tube filled with a rarefied
gas.  Horode\'{n}ski$^{(51)}$ concludes that more molecules will fly
parallel to the tube than across it.  This seems plausible, since
molecules moving parallel to the tube have longer free paths.
Therefore, if particles are allowed to migrate between the end of one
tube and middle of another, as in Fig.~3, there should be a net
migration from end to middle.  It follows that a clockwise wind is
expected in Fig.~3, which could operate a turbine.

Another example is sketched in Fig.~4. Two chambers are connected
through a
microscopic orifice, located at the extremity of a funnel-like surface. This
surface guides the molecules in the lower chamber to the orifice. On the other
hand, molecules in the upper chamber have a minute probability of escaping
through
the orifice. From this asymmetry we could expect a pressure difference to build
up, which could be used to extract work. Devices of this kind are successfully
used as traps for flies.

In the following examples an intrinsic asymmetry will be introduced to
the motion of the particles.  This may be achieved by means of magnetic
or Coriolis forces, which bring about a difference between turning to
the left or to the right.
\par
The first case is sketched in Fig.   5a. The ring represents a piece of metal
in
which the free electrons have very long free paths, so that they collide only
against
the boundaries of the ring. These electrons have thermal motion; in the absence
of
electromagnetic fields, they move along straight trajectories until they hit a
boundary of the ring. At the boundary, the electron velocity is randomized (in
accordance with the temperature of the ring) and a new straight trajectory
begins.
We classify the trajectories into two groups: trajectories in group I
contribute to a
clockwise current around the ring and trajectories in group II contribute to a
counterclockwise current (currents of electrons; the negative charge of the
electron is not essential for the present argument). On the average, both
currents
are equal and the net current is zero.
\par
Let us now consider the case in which the ring is embedded in a magnetic field,
perpendicular to the plane of the ring, such that the electrons move in
counterclockwise circular orbits. In this case trajectories in group I will be
curved opposite to the ring and will hit the boundary sooner than in the
zero-field case. Contrariwise, trajectories in group II bend in the same sense
as the boundary and will therefore have lengthened lives. As a result,
trajectories in group I are inhibited, in group II are enhanced and a non-zero
net current (which can run a motor) should be expected. While imparting
energy to the motor, electrons slow down.  Their energy is not
replenished by the static magnetic field, since the work performed by
the magnetic force is zero.   Rather, they pump thermal energy from the
environment (when they rebound at the walls), since its temperature
remains unchanged.

But Nature is stubborn. There are some trajectories, such as that shown
(enlarged)
in Fig.~5b, which contribute to a current in ``the wrong" sense when the
magnetic field is present. If the velocity distribution of the rebounding
electrons were such that there are many trajectories as in Fig.~5b and very few
grazing trajectories, then the entire effect could cancel out. This is indeed
the
case.$^{(53,54)}$ The distribution of the rebounding particles is
precisely the one
which is required in order that no net current builds up in Fig. 5, no
wind builds up in Fig.~3 and no
pressure difference builds up in Fig.~4. The latter device works for flies,
because they have prejudices, but molecules don't (as far as we know).

We shall close this subsection with a more sophisticated situation.  Let
us consider a gas of polarizable particles (as most particles are).  We
assume that the particles can be excited by some agent, say
electromagnetic radiation.  Let there be an electric field in the
$y$-direction and a magnetic field in the $z$-direction.

When a particle becomes excited, the interaction between its positive
and negative components becomes weaker and therefore its polarization
increases.  This means that the positive components will move in the
$y$-direction relative to the negative components, and the particle will
experience a net magnetic force in the $x$-direction.  As a result, when
a particle is excited it will be granted some momentum $\Delta p$ in
the $x$-direction, which is proportional to the magnetic field and to
the change in polarization.  When the particle is de-excited, the
change
in momentum is $-\Delta p$.  We might therefore expect a flow of excited
particles in the $x$-direction and a flow of de-excited particles in the
opposite direction, namely, heat flow  in the $x$-direction.$^{(55)}$

Preliminary simulations which I once performed did not provide evidence
for the expected heat flow.  Nevertheless, it seems to me that this
situation deserves further study.

Related ideas have been proposed by several authors.$^{(56)}$\\

\noindent{\bf 5. \ Typical and Exceptional Measurements}

As stated in Subsection 3.2, the entropy of the system may be
identified with the lack of information we have about the state of the
system.  The word ``we'' may disturb those who are used to think of the
entropy as a function of the system only; ``we'' are those who decide
which are the macroscopic coordinates, measure them and control them.
``Our'' role has been discussed in Section 3.
As a consequence, if we increase our knowledge by means of a
measurement, then the entropy of the system should decrease, even if the
microstate
of the system remains unchanged.  This possibility is what Smoluchowski
had in mind when he suspected that ``intelligent beings'' might overcome
the second law.
\par
The old answer to this apparent violation of the second law is that the entropy
of
the measuring instrument must increase by at least as much as the entropy of
the
system decreases. The question of whether it is or is not essential to invest
useful energy in
order to perform a measurement has been the subject of extensive
discussion.$^{(13,28,51,57-60)}$

For the purpose of the present section, we may disregard the question of
whether measurements could in principle reduce the entropy and consider
only the quantitative  aspect of this process.  How much could a
measurement help?  For instance, if we are dealing with a system of $n$
particles and in the ideal event that we exactly measure the state of
one particle, how much information do we gain?  The remaining lack of
information would be that of the remaining $(n-1)$  molecules.
Therefore, if $n$ is comparable to Avogadro's number, the decrease in
entropy would have no practical consequence.  More generally, what
measurements do is discard possibilities. For example, if before the
measurement there were
$N$ conceivable states $\psi_i,$ and the measurement adds three
significant digits to the previously known value of some quantity,
then, after the measurement there would
typically remain $N/1000$ possible states $\psi_j$ which could for instance be
those with $j=1000,2000,3000,...N.$ Let's denote by $H^\prime $ the value of
$H$
after the measurement. $H^\prime=\sum P^\prime_j\ln P^\prime_j$, with $P^\prime
_j$
the probability  for the system being in the state $\psi_j$, as inferred from
the
measurement. For simplicity, let's assume that if the system were known to be
before
the measurement in any of the states $\psi_i$ in the interval $j-999\leq i\leq
j,$
then it would be known to be in $\psi_j$ after the measurement. This means
\begin{eqnarray}
P_j^\prime=\sum^j_{i=j-999}P_i \ .     
\end{eqnarray}
It follows that the increase in $H$ due to the measurement is
\begin{eqnarray}
\Delta H=H^\prime -H&=& \sum_j\sum_{i=j-999}^jP_i\ln
P_j^\prime -\sum _j\sum
_{i=j-999}^jP_i\ln P_i \nonumber \\
&=&\sum_j\sum^j_{i=j-999}P_i\ln (P^\prime _j/P_i) \ .
\end{eqnarray}
This increase attains its maximal value when all the probabilities in every
group $j-999\leq i\leq j$ were equal before the measurement, as would
be the case if all the states in the group belonged to the same class
$G_{\nu(j)}$ of previously indistinguishable states. In this
case $P_i=P^\prime _j/1000$, and  we achieve an entropy decrease
$-\Delta S=k\Delta H_{\max }=k\ln 1000$.  Had the measurement provided
ten significant figures, then we would achieve $-\Delta S = 10 k\ln 10$,
which is still a microscopic quantity.

Since only macroscopic amounts of entropy are thermodynamically
significant, a macroscopic number of measurements is needed to have any
influence in this context.  We might still hope to perform all these
measurements, were it not for three strong limitations: (a) they have to
be reversible, otherwise useful energy would be dissipated; (b) they
have to be concluded before the state of the system changes and they
become irrelevant; and (c) they have to measure independent quantities.
These limitations rule out the possibility of long sequences of
measurements; it is necessary to perform macroscopic numbers of
measurements simultaneously.  But reversible measurements
involve$^{(28,58,59)}$ gradual motion of mechanical components or
gradual variation of fields, as well as permanent equilibration with
thermal reservoirs.  In summary, the decrease of entropy by means of
typical measurements requires control over a macroscopic number of
macroscopic devices.  Such a monster seems to be a subject for
philosophical preoccupation rather than a physically significant menace to the
second law.

The quantitative performance of a hypothetical Maxwellian demon has also
been considered by Leff.$^{(61)}$  From his results it can be concluded
that a single demon would have practically no effect; only a team of
demons might have significant influence.
\par
We still have to consider the case of ``exceptional'' measurements. For
example,
let us consider the situation proposed by Gabor.$^{(62)4}$  Fig.~6
shows a long cylinder of volume $V$ with a single molecule in its
interior.
Every time the molecule hits the walls of the cylinder,
it rebounds with a new
velocity. These velocities cannot be predicted deterministically, but they
comply with the
Maxwell-Boltzmann distribution corresponding to the temperature of the
cylinder
(which is the same as the temperature $T$ of the reservoir). Using kinetic
theory, it
is easy to show that the average force exerted by the molecule on a unit
surface
multiplied by the volume available to the molecule equals $kT$, so that
the average work performed by the molecule when moving a piston will be
a particular case of an ideal gas which changes its volume.

We assume that the molecule has some detectable feature, such as a magnetic
moment, which enables detection of the presence of the molecule within a
volume $v$. We could then place a detector at an end of the cylinder and wait
until the molecule is found within the volume $v$ close to that end.
This situation may be properly likened to a spider waiting for its prey. Once
the molecule is detected, it is imprisoned by the insertable piston $P$.
Enviewing the single molecule as  an ideal gas
which has changed its volume from $V$ to $v$, its entropy will
have decreased by the
amount $-\Delta S=k\ln (V/v)$. With the help of a mechanical device $W,$ our
single-molecule gas is now allowed to expand, while performing work on $W$. For
a
quasistatic, frictionless and isothermal expansion, the amount of performed
work
will be $kT\ln (V/v)$, with $T$ the prevailing temperature. In expanding, the
entropy of the gas returns to its previous value; but this time the entropy of
the
heat reservoir decreases by $k \ln  (V/v)$, violating the second law. By
choosing
$V$ sufficiently large, the gain in useful energy should more than compensate
for
any possible losses in the operation of this engine.
\par
However, increasing the volume $V$ is a gamble. If the detector $D$ is
at the non-zero temperature $T,$ it will sometimes be fooled by thermal
fluctuations
and wrongly believe that the molecule is there.
Therefore, there is always a  risk of
undergoing too many false alarms, and exhausting the useful energy in repairing
their consequences, before the molecule is caught.
If we choose a huge value for $V$,
then we should expect to wait a very long time until the molecule
happens to arrive
and this risk becomes certain failure.
\par
As a closer example to real life, we consider a case in which a physicist
receives
two barrels of water. Both barrels look the same and the physicist might pour
them
together into a bigger container, without suspecting that this process has any
significant influence on the entropy of the water. Let's assume, however, that
the
barrels were brought in an open truck; one of them stood under the sun most of
the
journey, and the other one was in the shade. Therefore, if the physicist
happens to
measure their temperatures, he or she will discover that they are not equal.
This
discovery enables him or her to extract a macroscopic  amount of useful
energy.
\par
At this point it is essential to notice that the hard stage in extracting the
above
useful energy was not in measuring the temperatures, but rather in guessing
that
there might be a temperature difference. It could just as well be the case that
both
barrels were at the same temperature but, say, the salt concentrations were
different. Translated into the situation of Fig.~6, an exceptionally
good guess would be to
locate the detector in the right place, so that the molecule arrives in a short
time in spite of the ratio $V/v$ being huge.  Measurements like these,
which  by conservative thinking are expected to produce a void answer,
but in exceptional situations could provide a macroscopic amount of
information, are what I call ``exceptional measurements''.

Finally, let me express exceptional measurements in the language of
algorithmic complexity.$^{(25-29)}$  Given a long string of digits, an
exceptional  measurement consists of finding (actually, guessing) two
simple algorithms which achieve the following: the first assigns a
weight to the position of every digit and the second predicts the values
of the digits with a success rate which, properly weighted by the first
algorithm, is appreciably larger than expected for random digits.  The
first algorithm defines the macroscopic coordinate which is being
measured, and the second expresses compressibility.  I am not aware of
any mathematical study of the size of the feasibility domain for this
process.\\

\noindent
{\bf 6. Interpretation}

The rebel's inquiry can be made to comply with the second law if we accept the
view
that ``order" is not a property of a system, but rather a property of the
observer
describing it.\\

\noindent{\bf 6.1 Obvious order and encoded order}

The universe is always in some definite state, which we shall
denote by $\phi $.
(The meaning of universe is an assembly of objects which interact only among
themselves.) Usually, $\phi $ is a complicated linear combination of the states
$\psi_i$, which were chosen according to the best of the observer's knowledge.
\par
The ``objective probability" of having the universe in its state $\phi $ is
always 1,
and the ``objective probability" for any other orthogonal state is always 0.
Therefore, according to definition  (5), the ``objective entropy" of
the universe is always zero.
\par
\par
Roughly speaking, we say that the state of a system
is ordered when we can recognize in it some intelligible pattern and we
can predict the evolution of the state with some microscopic detail.  A
state has ``encoded order'' if order can be found in it by looking at
the correct combination of coordinates.  Mathematically, encoded order
can be transformed into order by a suitable mapping.
Let us consider a few examples as an illustration of what I mean by
``encoded order''.\\
{\em a}.  Let us consider our solar system and an observer from outer
space which looks at us once every several years.  If there were no
gravitation and motion, the observer would always see us at the same
positions and say that our state is ordered.  But since there is
gravitation and motion, the positions of the planets at each observation
will apparently be unrelated to the previously observed positions and
the first conclusion should be that the solar system is undergoing
chaotic motion.  However, if the observer knows Kepler's laws, then the
observed positions will fit into orbits which describe the itinerary of
the planets before, between and after observations, and the order will
be as perfect as in the case of motionless planets.\\
{\em b}.  Any computer language deals with ``integers'' 1,2,3,..., which
have a natural order.  A normal person knows which is the succeeding
integer in the sequence.  This is plain order.  Most computer languages
can also generate ``random numbers'', such as 6835258, 9110669,
7652114,...  This sequence is actually a permutation of integers,
designed to avoid any correlation between the values of the numbers
and their order of appearance.  Namely -- complete disorder.  However,
someone who knows the algorithm and the seed from which the random
numbers are generated, can write a program which assigns the proper
ordinal to any value, or vice-versa.  Namely -- complete determinism.
For the algorithms used by most computer languages, the length of this
program would be quite small, and has nothing to do with the number of
permutations which can be performed among the integers.  (Discovering
the algorithm would be a different story.)\\
{\em c.} Consider $l$ ideal pendula
$(l>>1)$, with periods
$\tau
/(p+1),~\tau /(p+2),...\tau / (p+l)$,
where $p$ is some integer.
All the pendula are launched together, initially
in phase. After some time of the order of $\tau / l$, the pendula are
completely out of
phase. If the location of every pendulum is unrelated to its period, their
motion
looks chaotic. However, after time $\tau $, every pendulum will have undergone
an integer number of periods, and they will all be in phase.  One might
say that order was destroyed shortly after launching and created again
shortly before time $\tau$; however, if we number the pendula by their
period lengths and write down their equations of motion, it becomes more
sensible to assert that order is always present: it is entangled most of
the time and shows up every time $\tau$.\\
{\em d}.  Consider a sample with all its spins initially in the same
direction.
This alignment may be thought of as order.  If the spins
are perpendicular to the magnetic field, they precess around the
field axis.  Due to the inhomogeneity of the sample, each spin precesses
at a slightly different angular velocity and at a time $t_1$ the
angular distribution of the spins will have become isotropic in the
precession plane.  This isotropic distribution may be thought of as
disorder.  With an appropriate technique, the orientations of all the
spins at time $t_1$ can be inverted.  After this inversion, the
orientation of every spin is the same as if it had been precessing
backwards.  Therefore, at time $2t_1$, every spin will simultaneously
be back at its original orientation and the sample will have returned to
the original ordered situation.   We may therefore say that at time
$t_1$ there was  encoded order and that during the interval $t_1 \leq
t \leq 2t_1$ this order was decoded.  The effect described here is well
known and measurable; it is called ``spin echo''. \\
{\em e}.  When light from an object passes a piece of glass, its phase
is shifted.  If the glass is frosted, this phase-shift is a complicated
function of position, and the image of the object becomes
unrecognizable, as in Fig.~7a.  With an appropriate technique, these
phase-shifts can be reversed and the original image is recovered, as
shown in Fig~7b.$^{(63)}$
\par
Although we can always imagine a mapping which transforms encoded order
into readily observable order, it should not be understood from the
examples above that perfect decoding is physically attainable.  As in
the case of Maxwell's demons, decoders are subject to thermal vibrations
which limit the amount of order which can be recovered.  Thus, while a
mathematical computer is assumed to be capable of sorting any list of
numbers, a physical computer is expected to introduce mistakes if the
number of elements in the list is of the order of Avogadro's number;
likewise, while the technique in example ({\em e}) can restore features
of optical wavelength sizes, it cannot cope with features of atomic
size.  For an initially ordered state it is customary to define two
different relaxation times: a time $\tau_1$ after which the initial
order apparently fades away, and a time $\tau_2$ after which the initial
order cannot be recovered by means of the available decoding technique.

The concept of encoded order may seem to be incompatible with the
concept of algorithmic randomness.$^{(25-29)}$  Even if there exists a
mapping which transforms a complicated state $\phi$ into a simple state,
say $\psi_{100}$, the mapping may be as complicated as $\phi$, so that
handling it is as impractical as handling $\phi$.  We shall return to
this point in the following subsection.\\

\noindent{\bf 6.2 The source of useful energy}

Thermodynamics relates the useful energy to the entropy, and Section 3
expresses the entropy as a function of the statistical state $\{P_i\}$.
According to Section 3,
the probabilities  $\{P_i\}$ don't describe the objective
evolution of the
system, but rather the evolution of the knowledge which the observer has about
the
system. Obviously, an entity which doesn't feel the energy around it
and cannot decide what to do with that energy, has no useful energy
at its disposal.

The lesson called for is that the useful energy available from a system
depends on the information available about the state of the system.
The
meaning of Boltzmann's theorem is that, since the observer is unable to
follow in detail the evolution of the universe, his or her knowledge about the
universe (and therefore his/her ability to use its energy) will decrease as
time
elapses.
The second law does not imply that order turns into ``disorder". What happens
is
that readily observable order transforms into encoded order. If we knew how to
perform the appropriate decoding, entropy would remain constant.

But how can we always lose something which we don't have?  Before we
can lose information about our universe, we necessarily have to gain
it.   We might expect to gain information by means of measurements, but
Section 3 and all the literature on ``intelligent
demons''$^{(13,50,51,57-60)}$ indicate that ``typical measurements''
won't help.
An increase in our knowledge about the universe will occur if and only
if we
are lucky enough to perform an ``exceptional measurement", as
described in the previous section.

If we lived in a random universe, it would be utterly hopeless to trust on
``exceptional measurements" as our initial source of useful energy. But the
universe
is not random; on the contrary, it is always in some definite state $\phi $.
For any
given state $\phi $, the universe favours those creatures which are sensitive
to some
coordinates that  give rise to ``exceptional measurements". For example, when
we
decide whether to eat an apple or not, the number of parameters which we
really check is not very large; it could be roughly estimated by
dividing the cross section of an optic nerve by the cross section of an
axon.  In order to decide with certainty
whether that
apple is good or dangerous to our health, the number of parameters which we
ought to check is at least of the order of the number of nucleotides of
our genome, which is greater by roughly three orders of magnitude.
It is therefore a
miracle that we usually succeed. The only reason that nearly everything that
looks and
tastes like an apple is indeed an apple, is that we live in a non-random
environment,
and our senses are adapted to filter non-random coordinates.
\par
In ``exceptional measurements'', use is made not only of the information
provided by the measurement, but mainly of previous knowledge about the
prevailing order in the environment.  When we detect an oil field, the
amount of useful energy which we can control is proportional to the size
of the field, and is practically unrelated to Eq.\ (24).

Let us now confront the problem of algorithmic randomness.  The problem
is best stated in Ref.~29.  Let us assume that at some moment, say at
big-bang time, the universe was in a simple state $\phi_0$.  After some
time $\Delta t$, the universe will be in a not too different state,
$\phi_1 = {\cal U}\phi_0$.  In general, after time $\ell \Delta t$, the
state of the universe will be $\phi_\ell = {\cal U}^\ell\phi_0$.  Only
after a very long time $L\Delta t$, which is the Poincar\'{e} time, the
state of the universe will be $\phi_L\approx\Phi_0$ and will again
become simple.  For a general state $\phi_\ell$ it will be true that it
could be transformed, in principle, into $\phi_0$, but in order to
specify the required transformation we should stipulate the very long
number $\ell$.  Handling the number $\ell$ brings about a thermodynamic
burden which for most cases is effectively equivalent  to an additional
amount of entropy $\sim k\ln\ell\sim k\ln L$.

I claim that the limitation referred to in the previous paragraph is
overcome by successful observers.  The entire formalism of algorithmic
randomness is valid for ``universal computers'',$^{(25-27)}$ namely,
the length of the language which translates between two computers is
negligible in comparison with the length of the strings which they
handle.  Successful observers are not universal computers; they have
evolved hand in hand with their environment and have some number
$\ell_0$, such that $|\ell-\ell_0|<<\ell$, built in within their
``hardware''.  Therefore, successful observers have to find relatively
simple transformations, like ${\cal U}^{\ell-\ell_0}$, and, accordingly,
the environment does not look chaotic to them.
\par
When useful energy is dissipated, the order of the universe changes, and new
coordinates become non-random. As the order to the universe changes, new
creatures,
which are sensitive to these new coordinates, are more suited to succeed. The
rebel's
conclusion could rightly be ``I have useful energy (which enables me to think),
therefore I must have been seeing order in my environment''.\\

\noindent{\bf 6.3 Revision of thermodynamic  concepts.}

{}From Eq.~(1) it follows that
\begin{eqnarray}
{\partial S\over
\partial U}={1\over T};~~~~~~{\partial S\over \partial V}={p\over T}
\ ,
\end{eqnarray}
where $U,V,T$ and $p$ stand for the
internal energy, the volume, the temperature and the pressure, respectively.
\par
I have been claiming that $S$ depends on the complete set of
coordinates chosen by the observer.  But, can the temperature or the
pressure, which are directly measured, depend upon this choice?  By
definition, they could.  Eq.~(25) involves partial derivatives, and
their values depend on which arguments are kept constant in taking the
derivative.  When going from (1) to (25) it has been assumed that these
arguments are, besides $U$ or $V$, all the macroscopic coordinates in
the complete set such that varying them would mean doing work.

In order to have a precise idea of the significance of Eq.~(25) to the
measurement of $T$ and $p$, let us first deal with an artificially
simple model.  The conclusions will then be generalized.

We consider a system whose state is described by $f$ indices
$i_1,i_2,...i_f$,
where each index can take on the values $i_j=0,1,2,...$ The energy $E$
of the system is postulated to be
\begin{eqnarray}
E=\epsilon
\sum^f_{j=1}i_j \ ,         
\end{eqnarray}
where $\epsilon $ is a constant. This model is known as the Einstein model for
$f$ modes of
excitation.
\par
Let $E$ be the only macroscopic coordinate which is measured by the observer
$A$.
For simplicity, we shall assume that $A$ can measure $E$ exactly.
For a
given value of $E$, the number $\Omega_f(E)$ of possible microstates is
determined by
the recursion relation
\begin{eqnarray}
\Omega_l(E)=\sum^{E/\epsilon}_{i=0}\Omega_{l-1}(i\epsilon);~~~~\Omega_1(E)
=1 \ .   
\end{eqnarray}
For $E>>\epsilon$ (``classical" limit), (27) yields
\begin{eqnarray}
\Omega_f(E)={1\over (f-1)!}\left({E\over \epsilon}\right)^{f-1} \ . 
\end{eqnarray}
\par
Since the value of $E$ is the only information $A$ has about the state of the
system,
the probability for each microstate is just $1/\Omega_f(E)$ and the entropy is
\begin{eqnarray}
S_A(E)=k(f-1)\ln (E/\epsilon)-k\ln (f-1)!                    
\end{eqnarray}
Therefore, the temperature $T_A$ is given by
\begin{eqnarray}
1/T_A=dS_A/dE=k(f-1)/E             \ .            
\end{eqnarray}
\par
Let observer $B$ measure, in addition to $E$, the quantity
\begin{eqnarray}
D=\epsilon \sum^{f_D}_{j=1}i_j             \ ,    
\end{eqnarray}
with $0<f_D<f$. We have chosen to consider this quantity, since it is neither
independent of nor a function of $E$. For given values of $E$ and $D$, the
number of
possible microstates will be $\Omega_{f_D}(D) \Omega_{f-f_D}(E-D)$. Since there
is no
reason to associate to any of these states a preferred probability over the
others,
the entropy is
\begin{eqnarray}
S_B(E,D) & =&k(f-f_D-1)\ln [ (E-D)/\epsilon ] + k(f_D-1)\ln    
(D/\epsilon )\nonumber     \\
& - &k \ln [ (f-f_D-1)!(f_D-1)!]
\end{eqnarray}
and the temperature $T_B$ is given by
\begin{eqnarray}
1/T_B=(\partial S_B/\partial E)_D=k(f-f_D-1)/(E-D)      \ .  
\end{eqnarray}
\par
In general, $T_B\ne T_A$. Note that if $D$ maximizes $S_B(E,D)$ and $f>>1$,
then
$T_B\rightarrow T_A$, the reason being that the knowledge about $D$ is
just the same which could have been anticipated from the
knowledge of $E$. In this case, $B$ does not have more knowledge than
$A$.

For observer $B$ it would be more natural to regard the system as being
composite:
part of it comprises the first $f_D$ excitation modes, has energy $E_1=D$,
entropy
$k(f_D-1)\ln ( E_1/\epsilon (f_D-1)!)$ and temperature $T_1=E_1/k(f_D-1)$; the
second part comprises the remaining $f-f_D$ modes, has energy $E_2=E-D$,
entropy
$k(f-f_D-1)\ln ( E_2/\epsilon (f-f_D-1)!)$ and temperature
$T_2=E_2/k(f-f_D-1)=T_B$.
\par
But how does the thermometer know whether to measure $T_A,~~T_1$ or $T_2$? What
the
thermometer measures is its own energy, reached in equilibrium with the system,
under
well specified thermal contact. A thermometer designed to measure $T_1$ (resp.
$T_2$) should exchange energy with the first $f_D$ (resp. the last $f-f_D$)
excitation modes of the system, but not with the remaining modes. A thermometer
designed to measure $T_A$ should exchange energy with all the excitation modes.
Actually, what thermometer $A$ measures is the intermediate temperature
\begin{eqnarray}
T_{meas}=cT_1+(1-c)T_2           
\end{eqnarray}
where $c$ depends on the relative strength with which the thermometer is
coupled to
the excitation modes in both sets. If the average coupling per mode is the same
for
the first $f_D$ and for the last $f-f_D$ excitation modes, and considering
$f,f_D>>1,$ then $c=f_D/f$ and $T_{meas}=T_A$. If neither $T_1=T_2$, nor is the
coupling strength independent of the set to which an excitation mode belongs,
then
$T_{meas} \ne T_A$, and different thermometers could give different results.
This
could be a clue for observer $A$, indicating that an important feature about
the
state of the system is being overlooked.
\par
If $E_1$ and $E_2$ differ by at most a few orders of magnitude, whereas $f_D$
is
smaller than $f$ by {\it many} orders of magnitude, then $T_1$ will effectively
become infinity. From the practical point of view, an apparatus which exchanges
energy only with the first $f_D$ modes will not be called a thermometer or a
heat
bath, but rather a resonator. $T_1$ being infinity, the Carnot efficiency when
using $E_1$ to perform work would be 100\%. In this case, $E_1$ is called
mechanical energy, $E_2$ is called the internal energy and $T_2=(\partial
E_2/\partial S_B)_{E_1}$ is called the temperature of the system.

Let us now generalize the conclusions obtained from this model.  We
regard Eq.~(25) as the definition of temperature.
In order to fit this definition,
thermometers should be made so that they do exchange internal energy but are
impermeable to all the remaining macroscopic coordinates in the
complete set. For instance, let's consider atmospheric
temperature. An observer who is aware of evaporation, will always be careful to
distinguish between the temperatures read by a dry or by a wet thermometer. On
the
other hand, an observer who doesn't suspect that vapors exist, will at best
notice that
different thermometers show different temperatures due to some unknown reason,
and will
be forced to conclude that there is a transient irreproducible behaviour unless
saturation is reached.

For typical situations, the observer is able to classify
the microscopic degrees of freedom of the system into several groups, such that
the
interaction among different groups is weak. The possible criteria for this
classification are unlimited: location, chemical composition, kind of motion,
interaction with static fields, interaction with radiation, resonance
frequency, etc.,
and all possible combinations. Accordingly, an observer $A$ will define $a$
groups,
each with its own complete set of macroscopic coordinates, including
internal energy
$U_\alpha $, entropy $S_\alpha $ and temperature
\begin{eqnarray}
T_\alpha = {\partial U_\alpha \over \partial S_\alpha    
}~~~\alpha=1,2,...a  \ .
\end{eqnarray}
Similarly, an observer $B$ could define a different set of groups and a
different
set of temperatures
\begin{eqnarray}
T_\beta={\partial U_\beta \over \partial S_\beta }~~~\beta =1,2,...b \ .
\end{eqnarray}

All that has been said for temperatures can be translated to the case of
pressures.  The force per unit area exerted on an instrument depends not
only on the state of the system, but also on the instrument.  The higher
the permeability of the walls of the barometer to a kind of particles,
the lower the pressure which these particles will exert on them.  If the
properties of the different components of a mixture are known, then
barometer walls can be designed so that they are either completely
permeable  or completely impermeable to appropriate groups of
components; in this way different
barometers  add independent pieces of information.  There is no
essential difference between ``partial pressures'' and the temperatures
$T_\alpha$ of Eqs.~(35)-(36), except for the experimental fact that it
is very easy to vary the volume of all the groups by the same amount,
while it is very difficult to vary their energies by the same amount.

It is the ambiguity in the classification of the degrees of freedom
of the system, as manifested by Eqs.~(35) and (36),
which makes it possible to bring the rebel and Lord Kelvin to
terms. According to Kelvin's
formulation of the second law, if a closed system {\it is} at uniform
temperature (and all the other intensive parameters are uniform too), then {\it
it
is} impossible to transform part of its internal energy into mechanical
energy. The
present interpretation reformulates the second law in the form: ``If {\it an
observer
is unable to discover} in a closed system {\it any classification} such that at
least
two of the temperatures (or any other intensive parameter) in his/her set of
groups are
different, then {\it this observer} is unable to transform part of the internal
energy
of the system into mechanical energy."\\

\noindent{\bf Acknowledgements}

This article owes its present form to the careful revision of the
preliminary version by Reviewer C.  I am indebeted to A.W. Joshi for
permitting reproduction of material from Ref.~30, to the publishers of
Scientific American for permitting the use of Fig.~1, which was adapted
from Ref.~58 and to J. Feinberg for sending me Fig.~7.
\pagebreak

\begin{center}
{\bf References}
\end{center}

\begin{enumerate}
\item J.A. Wheeler and W.H. Zurek, eds. {\em Quantum Theory and
Measurement} (Princeton Univ., N.J., 1983).
\item J.S. Bell, {\em Speakable and unspeakable in quantum mechanics}
(Cambridge University Press, Cambridge, 1987).
\item K.G. Denbigh and J.S. Denbigh, {\em Entropy in Relation to
Incomplete Knowledge} (Cambridge University Press, Cambridge,
1985).
\item W.H. Zurek, ed., {\em Complexity, Entropy and the Physics of
Information} (Addison-Wesley, Redwood, U.S.A., 1990).
\item L. Boltzmann, Sitzungsberichte Akad.\ Wiss.\ (Vienna) part II,
{\bf 66}, 275 (1872), translated in Ref.~6.
\item S.G. Brush, {\em Kinetic Theory} (Pergamon, Oxford, Vol.~1 - 1965,
Vol.~2 - 1966, Vol.~3 - 1972).
\item P. Ehrenfest and T. Ehrenfest, {\em The Conceptual Foundations of
the Statistical Approach}, English Translation (Cornell University,
Ithaca, 1959).
\item R.C. Tolman, {\em The Principles of Statistical Mechanics} (Dover,
N.Y., 1979).
\item P.C.W. Davies, {\em The Physics of Time Asymmetry}, (University of
California, 1974).
\item M.C. Mackey, Rev.\ Mod.\ Phys.\ {\bf 61}, 981 (1989).
\item W. Shannon and W. Weaver, {\em The Mathematical Theory of
Communication} (Univ. of Illinois, 1949).
\item A.I. Khintchin, {\em Information Theory} (Dover, N.Y., 1957).
\item L. Brillouin, {\em Science and Information Theory}, 2nd ed.
(Academic, N.Y., 1962).
\item R.B. Ash, {\em Information Theory} (Wiley, N.Y., 1965).
\item R.W. Hamming, {\em Coding and Information Theory} (Prentice Hall,
N.J., 1986).
\item E.T. Jaynes, Phys.\ Rev.\ {\bf 106}, 620 (1957); {\bf 108}, 171
(1957).
\item  E.T. Jaynes in {\em The Maximum Entropy Formalism}, R.D. Levine
and M. Tribus, eds. (MIT Press, Cambridge Mass., 1978).
\item E.T. Jaynes, {\em Papers on probability, statistics and
statistical physics}, Synthese Library, Vol.~158 (R.D. Rosenkrantz,
ed.) (Reidel, Dordrecht, 1983).
\item E.T. Jaynes, in {\em Complex Systems -- Operational Approaches},
H. Haken, ed. (Springer-Verlag, Berlin, 1985).
\item A. Katz, {\em Principles of Statistical Mechanics} (Freeman, San
Francisco, 1967).
\item A. Hobson, {\em Concepts in Statistical Mechanics} (Gordon and
Breach, N.Y., 1971).
\item R. Balian, Y. Alhassid and H. Reinhardt, Phys.\ Rep.\ {\bf 131}, 1
(1986).
\item J.C. Maxwell, ``Diffusion'', {\em Encyclopedia Britannica}, 9th
ed. (N.Y., 1878), vol.~7, 220.
\item Course in Analytic Inventive Thinking, The Open University of
Israel, unpublished. A more accessible related reference is
G.S. Altshuller, {\em Creativity as an Exact Science} (Gordon and
Breach, N.Y., 1984).
\item R.J. Solomonoff, Inf.\ Control {\bf 7}, 1 (1964).
\item A.N. Kolmogorov, Inf.\ Transmission {\bf 1}, 3 (1965).
\item G.J. Chaitin, J.\ Assoc.\ Comput.\ Mach.\ {\bf 13}, 547 (1966).
\item C.H. Bennett, Int.\ J.\ Theor.\ Phys.\ {\bf 21}, 905 (1982).
\item W.H. Zurek, Phys.\ Rev.\ A {\bf 40}, 4731 (1989), and references
therein.
\item   J. Berger in {\em Horizons of Physics}, Vol.~2, N.\ Nath, ed.,
series editor A.W. Joshi (Wiley Eastern, New Delhi, 1994?).
\item J. von Neumann, {\em Mathematical Foundations of Quantum
Mechanics}
(Princeton Univ., 1955).  Relevant chapters reproduced in Ref.~1.
\item S.W. Hawking, Phys.\ Rev.\ D {\bf 13}, 191 (1976).
\item S. Frautschi, Science {\bf 217}, 593 (1982).
\item P.T. Landsberg, {\em The Enigma of Time} (Adam Hilger, Bristol,
1982).
\item O. Penrose, {\em Foundations of Statistical Mechanics}, (Pergamon,
Oxford, 1970).
\item O. Klein, Zeits.\ Phys.\ {\bf 72}, 767 (1931).
\item E.C.G. Stueckelberg, Helv.\ Phys.\ Acta {\bf 25}, 577 (1952).
\item E.T. Jaynes, Am.\ J.\ Phys.\ {\bf 33}, 391 (1965).
\item E.T. Jaynes, Phys.\ Rev.\ A. {\bf 4}, 747 (1971).
\item J. Berger, Eur.\ J.\ Phys.\ {\bf 11}, 155 (1990).
\item W. Pauli in {\em Sommerfeld Festschrift}, P. Debye, ed. (Hirzel,
Leipzig, 1928).
\item P.T. Landsberg, Phys.\ Rev.\ {\bf 96}, 1420 (1954).
\item J. Hurley, Am.\ J.\ Phys.\ {\bf 54}, 25 (1984).
\item R. Peierls, {\em Surprises in Theoretical Physics} (Princeton
Univ., N.J., 1979).
\item M. Smoluchowski, Phys.\ Z.\ {\bf 13}, 1069 (1912); M. Smoluchowski
in {\em Vortrage uber die Kinetische Theorie der Materie und der
Elektrizitat}, M. Planck, ed. (Teubner, Leipzig, 1914).
\item R.P. Feynman, R.B. Leighton and M. Sands, {\em The Feynman
Lectures on Physics} (Addison-Wesley, Reading, Mass., 1963), Vol.~1,
ch.~46.
\item L.G.M. Gordon, Found.\ Phys.\ {\bf 11}, 103 (1981) and {\bf 13},
989 (1983).
\item L. Brillouin, Phys.\ Rev.\ {\bf 78}, 627 (1950); R. McFee, Am.\
J.\ Phys.\ {\bf 39}, 814 (1971); R.E. Burgess, ed., {\em Fluctuation
Phenomena in Solids} (Academic, N.Y., 1965); J.C. Yater, Phys.\ Rev.\ A
{\bf 10}, 1361 (1974); E.P. EerNisse, Phys.\ Rev.\ A {\bf 18}, 767
(1978).
\item P.A. Skordos and W.H. Zurek, Am.\ J.\ Phys.\ {\bf 60}, 876 (1992).
\item H.S. Leff and A.F. Rex, Am.\ J.\ Phys.\ {\bf 58}, 201 (1990).
\item H.S. Leff and A.F. Rex, {\em Maxwell's Demon: Entropy,
Information, Computing} (Princeton Univ. and Adam Hilger, 1990).
\item A. Horode\'{n}ski, J. Vac.\ Sci.\ Technol.\ A {\bf 3}, 39 (1985);
Phys.\ Lett. A {\bf 122}, 295 (1987); J.\ Berger, J.\ Vac.\ Sci.\
Technol.\ A {\bf 5}, 382 (1987).
\item J. Berger, Am.\ J.\ Phys.\ {\bf 53}, 899 (1985).
\item H.P. Steinr\"{u}ck, K.D. Rendulic and A. Winkler, Surface Science
{\bf 154}, 99 (1985).
\item J. Berger, Solid State Com.\ {\bf 53}, 387 (1985).
\item H. Aspden, Nature {\bf 347}, 24 (1990).
\item L. Szilard, Z.\ Phys.\ {\bf 53}, 840 (1929), translated
in Ref.~1.
\item C.H. Bennett, Scientific American {\bf 257}, 88 (November
1987).
\item J. Berger, Int.\ J.\ Theor.\ Phys.\ {\bf 29}, 985 (1990).
\item C.M. Caves, W.G. Unruh, and W.H. Zurek, Phys.\ Rev.\ Lett.\ {\bf
65}, 1387 (1990).
\item H.S. Leff, Am.\ J.\ Phys.\ {\bf 58}, 135 (1990).
\item D. Gabor, Progress in Optics {\bf 1}, 111 (1964), reprinted in
Ref.~50. \item J. Feinberg, Optics Letters {\bf 7}, 486 (1982).
\end{enumerate}
\pagebreak

\begin{center}
{\bf Endnotes}
\end{center}
\begin{enumerate}
\item This idiom has been taken from Ref.~2.
\item In case of doubt about what is meant by inference or
probabilities, let me state that their meanings are the same as in
Ref.~21.
\item It is known, though, that Eq.~(2) is insufficient to predict
macroscopic evolution when the system undergoes an instability.  For
instance, if at time $t_o$ a ferromagnetic sphere slightly above its
Curie temperature is immersed in a cool fluid, then it is
experimentally found that some time later the sphere becomes magnetized
in a definite direction.  However, if at time $t_o$ there is perfect
spherical symmetry (to the best of our knowledge about the material and
about the present fields), then the evolution rule (2) is unable to
break that symmetry.
\item Landauer and Bennett have shown that the original analysis of this
problem was inadequate. See, for instance, R. Landauer in {\em Selected
Topics in Signal Processing}, S. Haykin, ed. (Prentice-Hall, Englewood
Cliffs,N.J.,1989).
\end{enumerate}
\pagebreak

\begin{center}
{\bf Figure Captions\\}
\end{center}
\begin{description}
\item{Fig.\ 1:} Two compartments separated by a door which opens when
kicked from the right, but not when kicked from the left.  The spring
closes the door after the molecule $M$ has passed through.
\item{Fig.\ 2:} A closed circuit, composed of a resistor, a diode and a
motor.
\item{Fig.\ 3:} Four long thin tubes, such that an orifice connects
between an end of each and the middle of its neighbour.  If more
molecules are moving parallel to the tubes than across them, then there
will be wind $A\rightarrow B\rightarrow C\rightarrow D\rightarrow A$.
\item{Fig.\ 4:} Two compartments separated by a funnel-shaped surface,
with a microscopic orifice in the middle.  The surface is intended to
``help'' the molecules reach the orifice when coming from below.
\item{Fig.\ 5:} Orbits described by free electrons inside a conducting
ring.  Their trajectories are bent by a magnetic field perpendicular to
this page. (a) I and II depict two possible orbits.  Presumably, these
are the typical cases.  (b) Another possible orbit, which must be taken
into account.
\item{Fig.\ 6:} A very long cylinder of volume $V$, with a small region
$v$ where the molecule can be detected and trapped.  M: molecule; D:
detector; P: insertable piston; T: heat reservoir; W: mechanical
device, tailored to respond to pressure on the piston and deliver
mechanical energy.
\item{Fig.\ 7a:} Image of a recognizable object, distorted by a frosted
glass.
\item{Fig.\ 7b:}  The image of Fig.\ 7a, restored by means of the
technique described in Ref.\ 63.
\end{description}

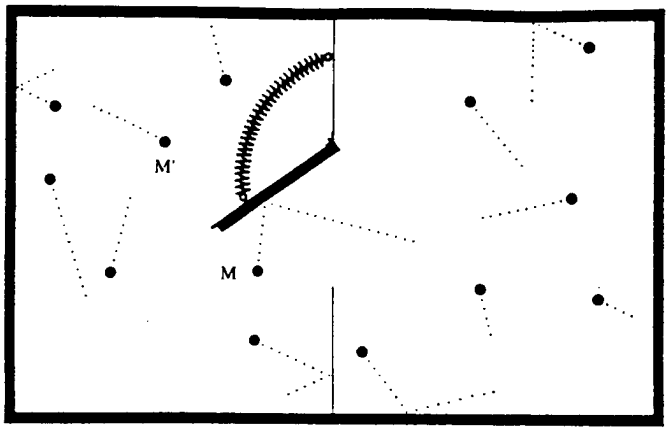
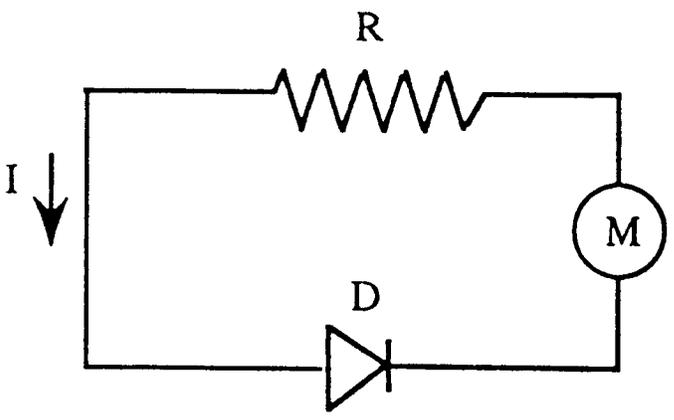
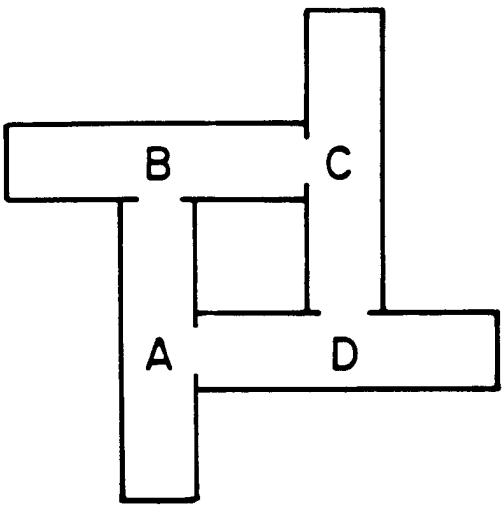
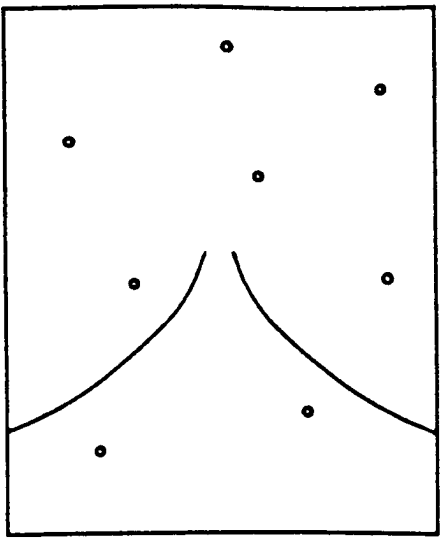
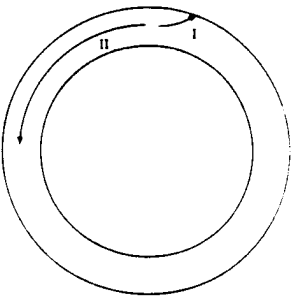
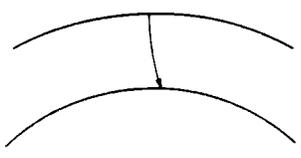
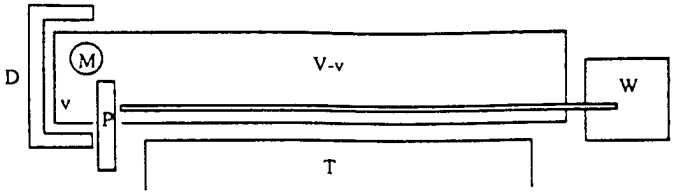
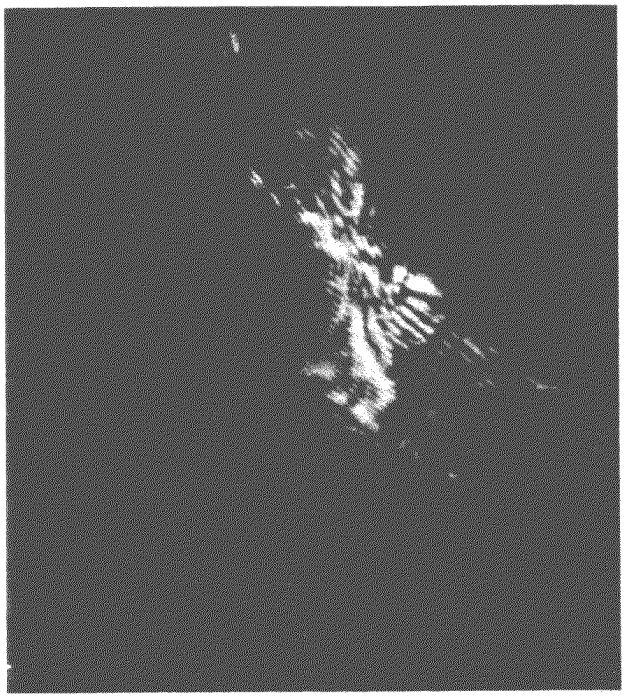
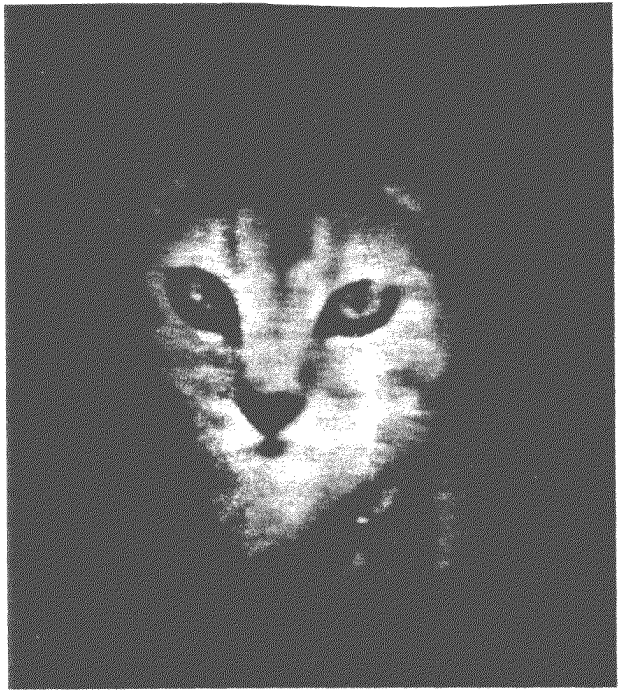
\end{document}